\documentclass[useAMS,usenatbib]{mn2e}
\usepackage{amssymb,amsmath} 
\usepackage{graphicx}
\usepackage{epstopdf}
\def \apj {ApJ\ }

\def \mnras {MNRAS\ }
\def \aap {A\&A\ }

\title[]{The {\it Swift} Gamma-Ray Burst redshift distribution: selection biases and optical brightness evolution at high-$z$?}

\author[]{D.M. Coward$^{1,2}$\thanks{E-mail:David.Coward@uwa.edu.au}, E.J. Howell$^{1}$, M. Branchesi$^{3,4}$, G. Stratta$^{5,6}$, D. Guetta$^{6,7}$, 
\newauthor B. Gendre$^{5,6}$, D. Macpherson$^{1,8}$, \\
$^{1}$School of Physics, University of Western Australia, Crawley WA 6009, Australia\\
$^{2}$Australian Research Council Future Fellow\\
$^{3}$DiSBeF - Universit\`a degli Studi di Urbino `Carlo Bo', I-61029 Urbino, Italy\\
$^{4}$LIGO - California Institute of Technology, Pasadena, CA 91125, USA \\
$^{5}$ASI Science Data Center, via Galileo Galilei, 00044 Frascati (RM), Italy\\
$^{6}$INAF - Osservatorio Astronomico di Roma, Via Frascati 33, I-00040 Monteporzio Catone (Roma), Italy\\
$^{7}$Department of Physics and Optical Engineering, ORT Braude, P.O. Box 78, Karmiel, Israel\\
$^{8}$University of Western Australia/International Centre for Radio Astronomy Research, M468, Crawley WA 6009, Australia\\
}

\begin{document}
\vspace{-5mm}

\pagerange{\pageref{firstpage}--\pageref{lastpage}} \pubyear{3002}

\maketitle

\label{firstpage}

\begin{abstract}
We employ realistic constraints on astrophysical and instrumental selection effects to model the Gamma-Ray Burst (GRB) redshift distribution using {\it Swift} triggered redshift samples acquired from optical afterglows (OA) and the TOUGH survey. Models for the Malmquist bias, redshift desert, and the fraction of afterglows missing because of host galaxy dust extinction, are used to show how the ``true'' GRB redshift distribution is distorted to its presently observed biased distribution. We also investigate another selection effect arising from a correlation between $E_{{\rm iso}}$ and $L_{{\rm opt}}$. The analysis, which accounts for the missing fraction of redshifts in the two data subsets, shows that a combination of selection effects (both instrumental and astrophysical) can describe the observed GRB redshift distribution. Furthermore, the observed distribution is compatible with a GRB rate evolution that tracks the global SFR, although the rate at high-$z$ cannot be constrained with confidence. Taking selection effects into account, it is not necessary to invoke high-energy GRB luminosity evolution with redshift to explain the observed GRB rate at high-$z$. 
\end{abstract}


\section{Introduction}
Long duration gamma ray bursts (GRBs) provide a potentially powerful probe of the early Universe. In particular, the spatial distribution of {\it Swift} satellite \citep{gehrels04} GRBs could be used to infer the star formation rate history at $z>2$. Although {\it Swift} has localized nearly 800 GRBs up to late 2011, only about 50\% of these have detected optical afterglows (OAs), despite rapid optical follow-up. Furthermore, only about 30\% have measured redshifts. Despite this deficit, there have been numerous attempts at analysing/simulating the GRB redshift distribution to constrain the star formation rate and/or GRB luminosity evolution at high-$z$ \citep[see e.g.][]{2005MNRAS.364L...8N,2006A&A...455..785M,2006MNRAS.372.1034D, 2008ApJ...683L...5Y, 2010MNRAS.406.1944W, 2010MNRAS.406..558Q,2011MNRAS.417.3025V}.

Previous work using the GRB redshift distribution as a constraint on rate/luminosity evolution \citep[e.g.][]{2006ApJ...642..371K, 2010MNRAS.406.1944W, 2012ApJ...749...68S} has not addressed the problem of the ``missing'' redshifts. Given that only $\sim30\%$ of {\it Swift} bursts have redshifts, the spatial distribution of the missing redshifts plays a fundamental role in understanding the link between GRBs, their environments, and relationship to the star formation rate. The first critical step in identifying where the missing redshifts are located is to understand why GRB optical afterglows are not observed for $\sim50\%$ of {\it Swift} bursts. In a previous study \cite{cow08} identified and constrained a global GRB redshift selection effect function, but used a purely empirical approach that did not include optical imaging limits. 

In this study, we use realistic constraints and models for redshift dependant selection biases, combined with GRB OA luminosities, to show how selection effects distort the ``trure'' spatial distribution to its presently observed distribution. We employ two subsets of GRB redshifts. The first, Howell \& Coward (2012), hereafter HC, uses 141 {\it Swift} triggered spectroscopic absorption redshifts from OAs up to Oct 2012. \footnote{GRB redshift sample is a subset taken from GCN circulars and http://www.mpe.mpg.de/$\sim$jcg/grbgen.html}. The second, less biased but smaller sample, uses a subset of 58 redshifts from the TOUGH (The Optically Unbiased GRB Host) survey \citep{2012ApJ...756..187H}. By accounting for selection effects, we investigate if the observed GRB redshift distribution is compatible with GRB rate evolution tracking the global star formation rate.  

\section{GRB optical selection effects}
\subsection{Instrumental selection effects}\label{SE1}
We define GRB OA selection effects as the combination of sensitivity limited optical follow-up and phenomena (astrophysical and environmental) that reduce the detection probability of an OA. Below we list the main instrumental biases which inhibit both OA detection and redshift measurement. These are discussed by \cite{2009ApJS..185..526F} and \cite{2012ApJ...756..187H} in more detail.

\begin{enumerate}\label{ biases}
\item {\bf XRT localisation:} An XRT localisation improves the probability of OA detection, especially in crowded fields.
\item {\bf Foreground extinction:} Galactic dust extinction significantly reduces the probability of detecting an OA.
\item {\bf Source declination:} The lack of rapid follow-up instruments capable of imaging at $\delta<\pm70$\degr. 
\item {\bf Source angular distance from Sun:} Bursts that occur too near the Sun have a limited follow-up window in time.
\end {enumerate}

Historically, because of the deficiency in pre-{\it Swift} ground-based follow-up of GRBs, there was a strong bias for imaging the brightest bursts. Because the brightest bursts are predominantly nearby, a significant faction of the first GRB redshifts were obtained by emission spectroscopy of the host galaxy. In both the pre-{\it Swift} and {\it Swift} era (from 2005 onwards), an optical afterglow (OA) is usually required to measure a redshift. For most high-$z$ GRBs, this is achieved by absorption spectroscopy of the GRB afterglow. The host galaxies are usually too faint to make a significant contribution to the spectra. Most GRB spectroscopic redshifts are acquired by ground based telescopes, including VLT, Gemini-S-N, Keck and Lick \citep[see][for a more complete list along with specific spectroscopy instruments]{2009ApJS..185..526F}.

The measurement of a GRB redshift depends strongly on the limiting sensitivity and spectral coverage of the spectroscopic system. For example, the VLT FORS 2 instrument has limiting magnitude of $\sim24$, while the VLT X-Shooter has a limit of $m\sim22$ for 3 x 60 min exposure times. One of the challenges for GRB spectroscopy is the fact that the OA is fading, so the delay time for acquiring spectra is also a factor. This bias is expected to manifest at high-$z$, where the brightest OAs are near the limiting sensitivity for spectroscopy. For faint OAs, photometric redshifts are measured by multi-band optical/NIR telescopes. For example, GROND \citep{2008PASP..120..405G} is providing photometric redshifts of high-$z$ GRBs via a seven channel simultaneous imager mounted on the 2.2 m MPI/ESO telescope at the ESO/LaSilla observatory. 

Other selection effects are more subtle, for example improving instrumentation and observation techniques over time that are used for both the OA follow-up and redshift measurement \citep{cow09}. The learning curve effect can be considered as a time-dependant bias because it affects the redshift distribution differently over time. An example is the slow drift in the median redshift during the first 4 years of {\it Swift} operation. Fig. \ref{zmean} shows the trend towards a smaller {\it Swift} mean redshift over the duration of the mission.\\
\begin{figure}
\centering
\includegraphics[scale=0.75]{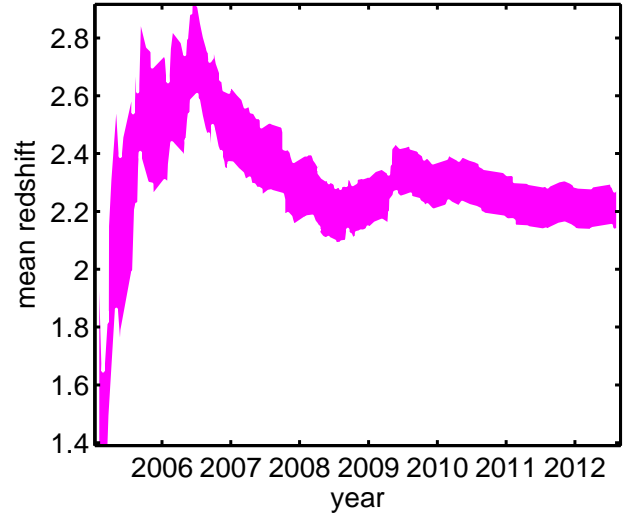}
\caption{The {\it Swift} mean redshift uncertainty bound plotted over the duration of the mission. It is clear there is a drift in the mean redshift over time, a consequence of different priorities and instruments contributing to redshift acquisition i.e. the learning curve effect \citep[see][]{cow09}. The {\it jump} observed in 2009 is a result of GRBs 090423 and 090429B, with redshifts of $z=8.26$ (NIR spectroscopic) and $z=9.2$ (photometric) respectively. } \label{zmean}
\label{zmean}
\end{figure}

\subsection{Redshift dependant selection effects}\label{SE2}
This following analysis focusses on the remaining selection effects that distort the redshift distribution over certain redshift ranges, with the assumption the GRB OA brightness is a proxy for redshift measurement i.e. a reasonably bright OA is usually required to obtain an absorption redshift.\\

\begin{enumerate}
\item {\bf Malmquist bias:} This bias arises because the telescopes and instruments acquiring OA absorption spectra (and photometry) are limited by sensitivity. In reality, the instruments acquiring redshifts are biased to sampling the bright end of the OA luminosity function. To account for this bias, it is necessary to have some knowledge of OA luminosity function (which is uncertain especially at the faint end), and an estimate of the average sensitivity limit of the instruments. This is the most fundamental bias that encompasses all peak flux limited detection and is the basis for modelling a selection function for OA/redshift measurement. We also note another selection effect for very high-$z$ GRBs. For very high-z GRBs the Malmquist bias is partially reduced by time dilation that extends the emissions implying it is easier to measure a redshift.

\item {\bf Redshift desert:} The so-called redshift desert is a region in redshift ($1.3 < z < 3$) where it is difficult to measure absorption and emission spectra. This becomes more significant for faint sources (small signal to noise ratio spectra). 
As redshift increases beyond $z\sim 1$, the main rest-frame optical emission features from the host galaxies (as for example the [OIII] 
$\lambda$3727 line, H$\beta$, H$\alpha$) become more difficult to detect since they move to the IR, a wavelength region where the sensitivity of CCDs starts to drop and sky brightness increases. At the same time, most of the strong UV absorption features in the OA spectra as those produced by the low ionization state metals (e.g. FeII, SII, SiII), and the weak rest-frame UV lines of active star forming galaxies are not redshifted enough to be detected less than $z=2-2.5$.

To demonstrate the effect of source brightness on number counts, Fig. \ref{QSO} shows the redshift distribution of faint quasars from the Sloan Digital Sky Survey Quasar Catalog (7th Data Release). For the faintest sources, (i.e. $m_R>20.9$) it is more difficult to acquire spectroscopic redshifts because these sources are most affected by the redshift desert, as shown by the strong `dip' in the redshift distribution.

\begin{figure}
\centering
\includegraphics[scale=0.65]{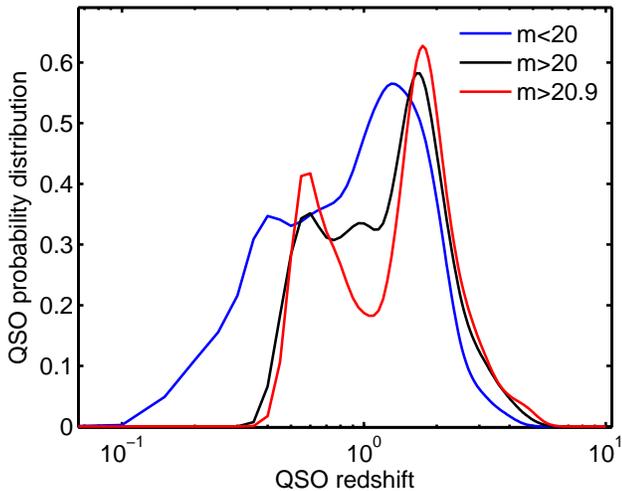}
\caption{The normalized redshift distribution of bright QSOs with $m_R<20$ (blue curve), faint QSOs $m_R>20$ (black curve), and very faint QSOs $m_R>20.9$ (red curve) from the Sloan Digital Sky Survey Quasar Catalog (7th Data Release). It is clear that the selection of fainter QSOs results in increasing selection effects. It is more difficult to acquire redshifts in the redshift desert ($0.8<z<2$) for the faintest sources. } \label{QSO}
\end{figure}

It is difficult to quantify the fraction of missed desert redshifts across different surveys because of the varying strategies and instruments used. For example, \citet{2005mmgf.conf..169S} find a typical success rate of (70-80)\% for obtaining spectroscopic redshifts of photometrically selected galaxies in the desert. This success rate is expected to improve for galaxies in $z = 2-2.5$, which are more suited for follow-up spectroscopy in the near-IR. 

\item {\bf Host galaxy extinction:} \label{dustitem}
 There has been growing evidence that dark bursts are obscured in their host galaxies \citep[e.g.][]{jakob04,2006ApJ...647..471L,2009AJ....138.1690P,2012MNRAS.421...25S}. The detection of the near-IR OAs of some GRBs (which would have been considered as dark bursts because their OAs were not detected in any bluer bands) provides evidence for dust obscuration (e.g. Tanvir et al. 2008). These studies generally show that GRBs originating in very red host galaxies always show some evidence of dust extinction in their afterglows.\\ 
 
 \citet{2009AJ....138.1690P} show that a significant fraction of dark burst hosts have extinction columns with $A_V \sim1$ mag, and some as high as $A_V =2-6$ mag. \cite{2012A&A...545A..77R} performed a search for the host galaxies of 17 bursts with no optical afterglow. They find in seven cases extremely red objects in the error circles, at least four of them might be dust-enshrouded galaxies. The most extreme case is the host of GRB 080207, one of the reddest galaxies ever associated with a GRB. Using multi-wavelength observations of GRB 080207, \cite{2012MNRAS.421...25S} show that that the dark GRB hosts are systematically more massive than those hosting optically bright events, perhaps implying that previous host samples are severely biased by the exclusion of dark events.\\

\citet{2011ApJ...735....2Z} compared the high redshift GRBs to a sample of lower redshift GRB extinctions and found a lack of even moderately extinguished events ($A_V\sim0.3$) at $z >4$. \cite{2009ApJ...705..936B} studied the effect of dust obscuration on estimates for the high-$z$ star formation rate. Their work showed that star-forming galaxies at $z >5$ almost universally have very blue UV-continuum slopes, and that there are not likely to be a substantial number of dust-obscured galaxies at $z > 5$ that are missed in ``dropout'' searches.
In spite of the biased selection and small number statistics, their analysis supports the argument for a decrease in dust content in star-forming environments at high redshifts. Hence it is highly likely that dust obscuration in GRB host galaxies is linked to the evolving star formation rate. This implies that the missing redshifts (with obscured OAs) may be tracking the star formation rate of red dusty star forming galaxies, while the observed redshift distribution is tracking the star formation rate of bluer more compact star forming galaxies. Supporting the high-dust at small redshift scenario, \cite{2012ApJ...754...89W} show that GRBs with high X-ray absorption column densities found at $z<4$ typically have very high dust extinction column densities, while those found at the highest redshifts do not. 

The most significant issue with interpreting a lack of highly extinguished OAs at high-$z$ as evidence for less dust extinction, is selection effects. Specifically, the non-obscured, bright OAs are preferentially selected at high-$z$, so that a possible obscured subset of GRBs remains hidden.  
Our independent approach to constrain an upper limit on GRB optical dropouts from host galaxy dust with redshift is based on optical surveys of core-collapse supernovae (SNe). \cite{2012ApJ...756..111M} find for a volume-limited rest-frame optical SN survey, the missing SN fraction increases from its average local value of about 19\% to 38\% at $z =1.2$, and then decreases marginally beyond $z = 2$. \cite{2012ApJ...756..111M} note that their estimated missing SNe fraction is also consistent with the recent findings of \cite{2012ApJ...755...85M} for their sample of $z <1$ GRB host galaxies. 

\end{enumerate} 

\begin{figure}
\centering
\includegraphics[scale=0.55]{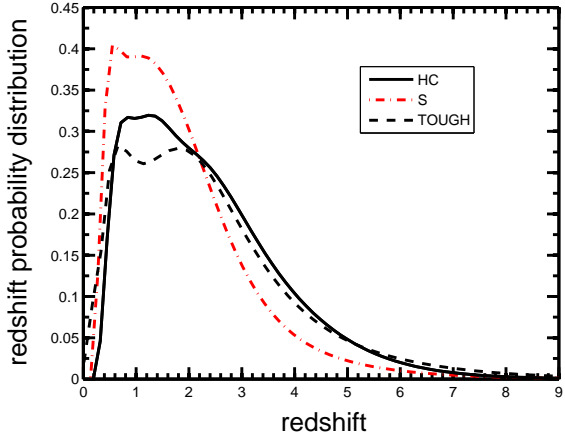}
\caption{A statistical comparison of three sub-sets of GRB redshifts: from \citet{2012ApJ...749...68S} -- (S), \citet{2012ApJ...756..187H} -- TOUGH and the subset of \citet{2013MNRAS.428..167H} -- (HC), plotted in Figure \ref{reds}. The HC sample has 3 times as many bursts, and is a very similar distribution to that of TOUGH: P$_{\mathrm{KS}}$=94\%. Alternatively the S sample, with a peak flux cut-off of 2.6 ph s$^-1$, is biased against high-$z$ redshifts. } \label{reds2}
\end{figure}

\subsection{Data selection}

Our primary redshift sample is based on the Howell and Coward (2012) subset from the J. Greiner catalogue of well-localized GRBs consisting of 141 {\it Swift} triggered spectroscopic absorption redshifts from OAs up to Oct 2012. We note that the HC redshifts are mostly measured from the OA, as opposed to the host galaxy, so that OA brightness is the key factor in determining a redshift. We also include photometric redshifts with $z>6$, because of the difficulty for obtaining spectroscopic redshifts at very high-$z$ because of Lyman-$\alpha$ dropout of optical sources.

We also compare two other catalogues of high redshift sub-samples of {\it Swift} long GRBs: the TOUGH sample and the sample of \cite{2012ApJ...749...68S}, which we label as S. For comparison we show in Figure \ref{reds2} the HC sample (141) with the samples of TOUGH (53 bursts) and S (51 bursts): we note that in this latter sample 4 bursts (GRB050802, GRB060306, GRB060614, GRB081221) have uncertain redshifts according to the HC sample and an additional burst, 050416A, is catagorised as an XRF. Figure \ref{reds2} shows that the the HC sample has a very similar distribution to that of TOUGH, i.e. a two sample P$_{\mathrm{KS}}$=94\%. The S sample shows that a peak flux cutoff at 2.6 ph s$^-1$ has produced a significant deficiency of bursts above $z \sim 2$ in comparison with the TOUGH and HC samples. This is supported by the corresponding two sample P$_{\mathrm{KS}}$ probabilities of 9\% and 2\% respectively. We find that by applying similar peak flux cut in our sample (reducing our sample to 52 bursts) our distribution resembles that of S, yielding a P$_{\mathrm{KS}}$ probability of 83\% (and 45\% when compared to the TOUGH sample). 

We choose to employ the HC subset because it:  a) demonstrates high statistical compatibility with the TOUGH sample; b) allows us to use a larger sample (nearly 3 fold number increase compared to the TOUGH and S samples). We conclude that the technique of removing the faintest GRB peak fluxes, sample S, introduces a redshift dependant bias that manifests as a deficiency of high-$z$ redshifts, as shown in Figure \ref{reds2}.

\begin{figure}
\centering
\includegraphics[scale=0.6]{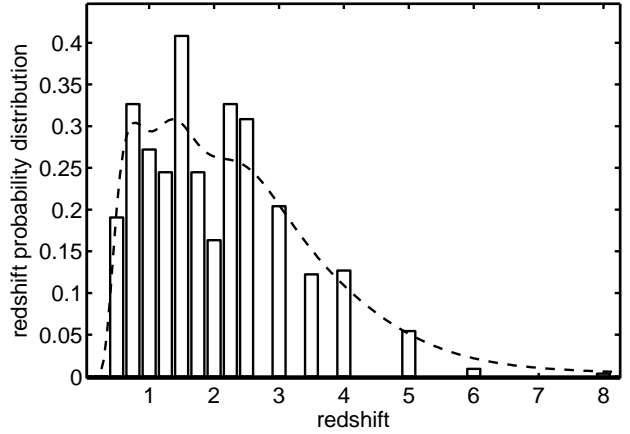}
\includegraphics[scale=0.6]{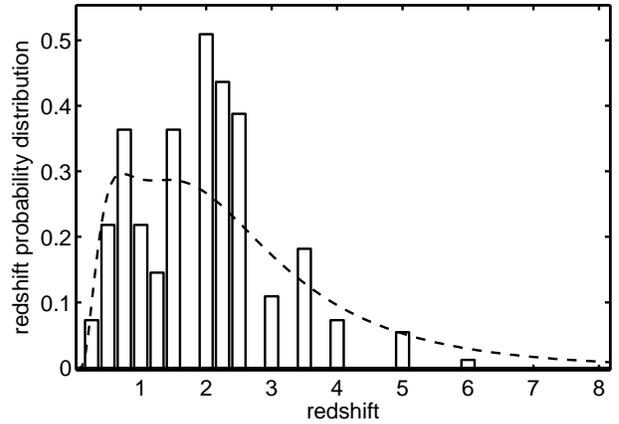}
\caption{{\bf Top:} The redshift distribution of 147 {\it Swift} GRBs (up Oct 2012) acquired by absorption spectroscopy of the optical afterglow, the HC sample, (normalized bars), but including six $z>6$ NIR spectroscopic and photometric redshifts, along with a number density curve (dashed line) that smoothes small scale fluctuations. {\bf Bottom:} The TOUGH survey subset of 58 redshifts obtained using strict selection criteria to minimize observational biases.} \label{reds}
\end{figure}

The HC redshift subset is representative of about 37\% of the total {\it Swift} sample. The TOUGH survey subset of 58 redshifts, obtained using strict selection criteria to minimize observational biases, is 55\% complete in redshift. This means that the two data sets require different normalizations to account for the missing redshift fractions. For the TOUGH data set, we assume that the instrumental biases are mostly accounted for, so that the remaining 45\% of redshifts are missing because of the three redshift dependant biases, namely the redshift desert, host galaxy dust obscuration and the Malmquist bias. For the HC subset, the normalization is 18\% missed because of instrumental biases, and 55\% missed because of the redshift desert, host galaxy dust obscuration and the Malmquist bias. Figure \ref{reds} plots the HC and TOUGH redshift distribution along with their density approximations. 

\section{GRB redshift distribution model with selection effects}\label{prob}
 
The dominant GRB redshift distribution biases discussed above are represented as the product of independent dimensionless selection functions that are unity for a 100\% selection probability: 
\begin{enumerate}
\item $\psi_{\rm Obs}$ -- number dropouts from mostly non-redshift dependant biases listed in \S \ref{SE1}, which are different depending on the selection criteria for the sample. We assume that the TOUGH sample is relatively free of instrumental biases, but about $20\%$ of the HC sample is affected by instrumental biases.     
\item $\psi_{\rm Swift}(z)$ -- the limited sensitivity of {\it Swift} to trigger on GRBs (see Appendix \ref{app} for a derivation of $\psi_{\rm Swift}(z)$).
\item  $\psi_{\rm M}(z)$ -- the limited sensitivity of instruments to measure a redshift from the GRB OA.
\item $\psi_{\rm Desert}(z)$ -- number dropouts from the redshift desert.
\item $\psi_{\rm Dust}(z)$ -- number dropouts from host galaxy dust extinction. 
\end{enumerate}

\begin{figure}
\centering
\includegraphics[scale=0.35]{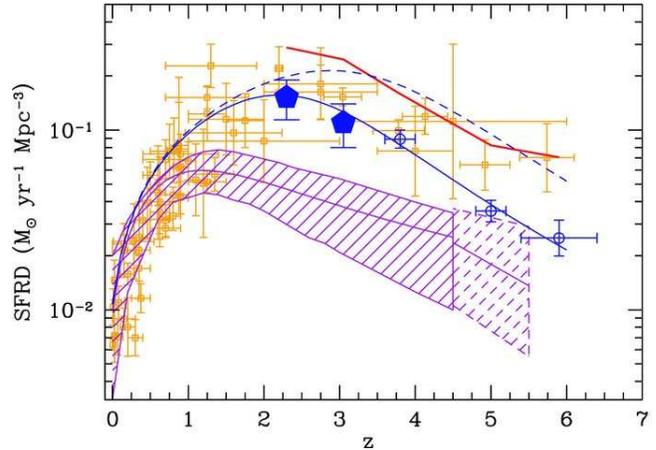} 
\caption{{\bf Top panel}: Star formation rate history from multi-wavelength surveys taken from Fig. 10 in \citet{2009ApJ...692..778R} and reproduced by permission of the AAS. We use the solid line, for the best-fit star-formation history assuming a luminosity-dependent dust correction to $z\approx2$. See Appendix \ref{ez} for the conversion between the SFR and GRB rate evolution model, $e(z)$.} \label{sfrfig}
\end{figure}

The GRB redshift probability distribution function, that includes the above selection effects, can be expressed as: 
\begin{equation}
\begin{array}{ll}
P(z) = & N_p\frac{dV(z)}{dz}\frac{e(z)}{(1+z)} \psi_{\rm Swift}(z)\psi_{\rm Obs} \psi_{\rm M}(z)\\
&  \psi_{\rm Desert}(z) \psi_{\rm Dust}(z)
\end{array}
\end{equation}
where $N$ is a normalization constant. The volume element, $dV/dz$, is calculated using a flat-$\Lambda$ cosmology with $H_{\mathrm 0}$ = 71 km s$^{-1}$ Mpc$^{-1}$, $\Omega_M$ = 0.3 and $\Omega_\Lambda$ = 0.7, and we fix $\psi_{\rm Obs}\approx0.5$ (see the selection effects listed in \S \ref{SE1}). The function $e(z)$ is the dimensionless source rate density evolution function (scaled so that $e(0)=1$). We assume that $e(z)$ tracks the star formation rate history, and we employ the star formation rate best fit to multi-wavelength surveys from Fig. 10 in \cite{2009ApJ...692..778R}, shown in Fig. \ref{sfrfig}. Because the fit only extends to $z=6$, we extrapolate the fit to $z=10$. Fig. \ref{ezfig} plots $e(z)$ using a simple functional form: two linear increases from $z=0-2.5$, and an exponential decrease at $z>2.5$ (see Appendix \ref{ez}).

The question of how the GRB rate evolves with metallicity remains controversial. \citet{Elliott2012} found no evidence for GRB rate evolution tracking metallicity from a sub-sample of GROND, and that various redshift biases \citep{cow08} could account for previous interpretations. Because of this unceratinty, we adopt this argument and do not include metallicity evolution, but instead assume GRBs track the global evolving star formation rate, i.e. $e(z)$ above.

The total selection function, i.e. excluding $\psi_{\rm Swift}(z)$, is defined as:
\begin{equation}\label{frac}
\psi_{\rm Total}=\psi_{\rm Obs}\times\psi_{\rm M}(z)\times \psi_{\rm Desert}(z)\times\psi_{\rm Dust}(z)\;.
\end {equation}
The total fraction of GRB redshifts that are missed because of optical selection effects can be expressed as
\begin{equation}
F_T=\bigg[1-\frac{N_p^{-1}\int_z^{10}P(z) \mathrm{dz}}{\int_z^{10} dV(z)/dz (1+z)^{-1} e(z) \psi_{\rm Swift}(z) \mathrm{dz}}\bigg]\;,
\end {equation}
where the normalization is cancelled to convert to a relative number density. This provides an independent consistency check of $\psi_{\rm Total}$, because $F_T\approx 63\%$ (45\%) for HC and TOUGH respectively, is based on the fraction of measured redshifts to the total number of {\it Swift} bursts. We apply the Kolmogorov-Smirnov test (KS test) test using the model $P(z)$ that includes selection effects, and the two data sets  HC and TOUGH. In addition to the KS test we employ an error calculation of the fraction of redshifts missed in the model compared to the observed missed fraction. The fractional error, $E$ is:

\begin{equation}
E = (F_T+I-N)/N,
\end {equation}
where $I$ is the fraction missed because of instrumental biases, which we assume is zero for TOUGH and 0.18 for the HC sample. The normalization constant, $N$, for HC and TOUGH is 0.63 and 0.45 respectively.

\subsection{Optical afterglow selection function}
The GRB OA luminosity function (LF) in R-band is approximated by fitting to the compiled OA luminosities from \cite{2010ApJ...720.1513K}. Their luminosities, scaled to 1 day, have been corrected for Galactic extinction, host extinction and are k-corrected if possible. We use the functional form of \citet{Johannesson2007} to approximate the LF: 
\begin{equation}
\label{eqn:LF}
\varphi(L) = C\left(\frac{L}{L_0}\right)^{-\lambda}\exp\left(-\frac{\ln^2(L/L_0)}{2\sigma^2}\right)\exp\left(-\frac{L}{L_0}\right)\;,
\end{equation}
where $C$ is a normalization constant, $L_0$ a characteristic luminosity. We convert $L/L_0$ to absolute magnitudes using $L/L_0=10^{(M_0 - M)/2.5}$. Using the KS test as a constraint ($P_{\rm{KS}}>0.8$), we approximate the LF using $M_0=24.5$, $\sigma=3.5$ and $\lambda=0.5$. This LF was converted into a probability distribution function (PDF), plotted in Fig. \ref{Kann}, by normalizing over the luminosity range $[-28,-16]$, where we have extended the faint end to $M=-16$ to crudely account for the dropout of faint OAs from the Malmquist bias. 

We note that the above LF does not take into account the Lyman-$\alpha$ cut-off at high-z R-band sources. Lyman-$\alpha$ absorption renders sources at $z>6$ undetectable in the R-band. Nonetheless, the GRB redshift data contains 6 sources with either spectroscopic or photometric redshifts at $z>6$, which are significant in our analysis. For these very high-$z$ GRBs, we assume that the OA brightness in filters redder than R-band (including the near-infrared) follows approximately the same number distribution as in R-band (although in the H-filter the GRB OAs will be brighter from 1-2 mags but the CCD sensitivity reduces by a similar magnitude).

\subsubsection{A Malmquist bias with a $E_{{\rm iso}}-L_{{\rm opt}}$ correlation}\label{corr}

In a study of both short and long duration GRBs \citet{2008ApJ...689.1161G} found optical, X-ray and $\gamma$-ray emissions to be linearly correlated. A highly significant correlation $( > 99.99\% )$ was found between the prompt $\gamma$-ray fluence and X-ray flux of long duration GRBs (r $\sim 0.53$ ). A slightly less significant correlation $( 99\%)$ was revealed between the optical and X-ray fluxes (correlation coefficient, $r \sim$ 0.44). Similar correlations existed for the short duration GRB sample, although were less significant due to the smaller sample. Stronger bursts in the X-ray and $\gamma$-ray produced brighter afterglows in the optical and X-ray respectively. A large spread was seen around the correlation line in all instances.

\cite{2010ApJ...720.1513K} find a weak correlation between $E_{{\rm iso}}$ and $L_{{\rm opt}}$ with a Kendall rank correlation coefficient of 0.29 for the subset of GRBs used in this study to estimate an optical LF. The key question is: how does this correlation affect the Malmquist bias? Firstly consider the effect if $E_{{\rm iso}}$ and $L_{{\rm opt}}$ are uncorrelated. This implies that a high $E_{{\rm iso}}$ (that will be preferentially detected by {\it Swift}) could be associated with equal probability with either a low or high luminosity optical afterglow. This scenario implies that detection of the OA is independent of the high energy luminosity. Alternatively, observation suggests a high $E_{{\rm iso}}$ preferentially selects a high optical luminosity. This implies that the Malmquist bias will be reduced at high-$z$, because it is the high $E_{{\rm iso}}$ bursts that are preferentially seen at large-$z$. These bursts will also be more optically luminous, so that redshift measurement will be more probable.

Figure \ref{corr} plots the 1-day optical luminosities (at $z$=1) \citep{2010ApJ...720.1513K} (79 bursts) against bolometric isotropic burst energy. We also plot the sub-sample used in this study (from the HC sample) - obtaining a subset of 48 bursts. On the same Figure we calculate the change in mean optical luminosity across $E_{{\rm iso}}$ by binning the data. It is clear that there is a trend for increasing mean OA luminosity with increasing $E_{{\rm iso}}$. The change in magnitude, $\Delta M$, across is $E_{{\rm iso}}$ is about $\Delta M\approx1.9$. As described above, the correlation will reduce the Malmquist bias at high-$z$. We employ the constraint of a maximum increase in luminosity of $\Delta M\approx1.4$ mag in $z=0-6$. The Mamquist correction is defined as $\Delta M(z) = 1.7\rm{Log}_{10}(1+z)$, so that $\Delta M(6)\approx1.4$.

\begin{figure}
\centering
\includegraphics[scale=0.5]{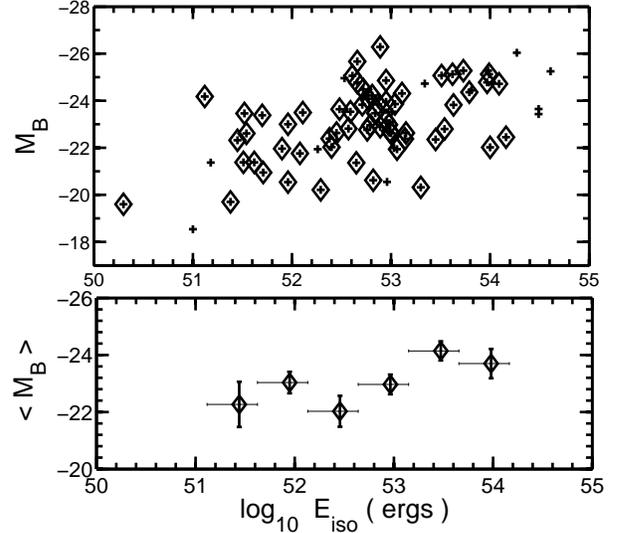}
\caption{{\bf Top panel}: The 1-day luminosities (at $z$=1) for the Kann sample of optical afterglows (79 bursts) are shown as '+' symbols against bolometric isotropic burst energy. The sub-sample used in this study - 48 bursts obtained via spectroscopic absorption and designated as LGRBs in the HC catalogue are indicated by diamonds. We find a weak correlation with a Kendall rank correlation coefficient of 0.21 (with 97\% significance) - this is less than 0.29 obtained by Kann et al (2010) for their full optical luminosity sample. {\bf Bottom panel}: The mean magnitudes and their 1-$\sigma$ errors obtained per logarithmically spaced bins of equal width. We use 7 bins (indicated by horizontal the error bars) which is the maximum number that obeys a requirement of at least 5 events per bin. The drift in the mean magnitude with increasing $E_{{\rm iso}}$ of the sample, $\Delta M$, is about 1.4 mag.} \label{corr}
\end{figure} 

\begin{figure}
\centering
\includegraphics[scale=0.75]{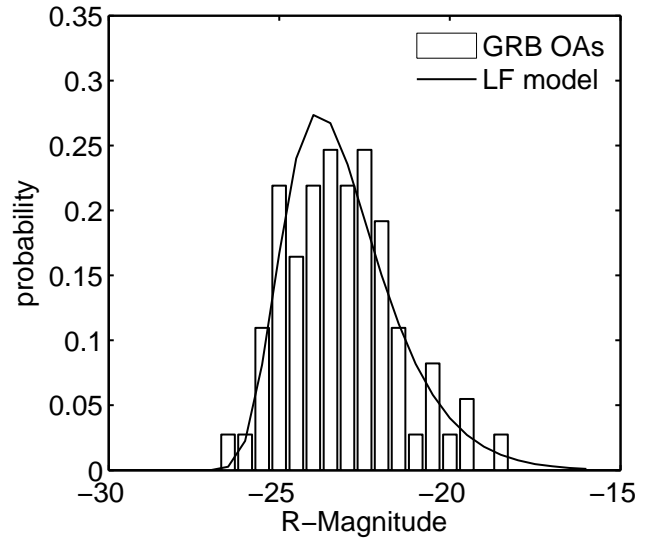}
\caption{The GRB OA probability distribution function (PDF) using equation \ref{eqn:LF} and GRB OA luminosities from Kann et al. (2010). The optimal model is constrained by the KS test ($P_{\rm{KS}}>0.8$) with $M_0=24.5$, $\sigma=3.5$ and $\lambda=0.5$, and normalized over the Magnitude range $[-28,-16]$. We note that the faint end is the least constrained because of the Malmquist bias.} \label{Kann}
\end{figure} 

\subsubsection{Malmquist bias}\label{malm}
A telescope limiting magnitude ($m_\mathrm{L}$) defines a threshold for obtaining a redshift. Because of the different instruments engaged in spectroscopic redshift measurement it is difficult to constrain this important parameter. For definiteness, we take $m_\mathrm{L} = 24$, approximating an average limit for the telescopes used to acquire the redshifts for the faintest bursts. We include an average extinction from Kann et al. (2010) of $A_V=0.3$.   

With a reference time $t_\mathrm{c}$ of 1 day and assuming that all OAs fade in luminosity by $t^{-1}$, the OA limiting luminosity was calculated at a time $T_\mathrm{z}$ \citep{imerit09} as:
\begin{equation}
\label{eqn:Pogson}
M_\mathrm{L}(z) = m_\mathrm{L} - 5\log_{10}\Big(\frac{d_\mathrm{L}}{10}\Big) - \frac{5}{2} \log_{10}\Big(\frac{T_\mathrm{z}}{t_\mathrm{c}}\Big) + k(z) + \Delta M(z)-A_V.
\end{equation}
For the mean time taken to acquire a redshift \citep{cow09}, we use an average $T_\mathrm{z}\approx400$ min. Hence $T_\mathrm{z}$ links the probability of redshift measurement to the brightness of the OA. For the k-correction we use 
$-2.5 (\beta-1){\rm Log}_{10}[1+z]$, with $\beta=0.66$ from \cite{2010ApJ...720.1513K}. We employ two scenarios for the limiting magnitude. The first assumes a correction for the correlation of higher $E_{{\rm iso}}$ with brighter optical luminosities, $\Delta M(z)$ (see section \ref{corr}). We also calculate the Malmquist bias without this correction.
Using the above definitions, the OA selection function defines the fraction of observable OAs with redshift:
\begin{equation}
\label{select2}
\psi_{\rm OA}(z) = \int_{M_\mathrm{L}(z)}^{M_\mathrm{Max}}{\varphi(M)} \mathrm{dz}\;.
\end{equation}

The total fraction of GRB redshifts missed because of the Malmquist bias is less than 10\% using the above model assumptions. 

\subsection{OA host galaxy dust obscuration}
For an estimate of optical GRB number dropouts from host galaxy dust, we employ the study of \cite{2012ApJ...756..111M} using a volume-limited rest-frame core-collapse SN survey. Their missing SNe fraction is obtained in two parts: First, they estimate the fraction of SNe in normal galaxies with substantially higher host galaxy extinctions than predicted by simple models for the smooth dust distribution and the resulting inclination effects. 

Secondly, they estimate the fraction of SNe missed in local U/LIRGs. For this they make use of the rich SN population of one of the nearest LIRGs, Arp 299. Assuming that the SNe with the highest host galaxy extinctions are missed by the optical searches and not compensated for by the standard extinction corrections, they derive the fraction of SNe that remain missing and estimate the corrections needed to be applied when deriving CCSN rates. Finally they use this information together with the latest knowledge of the nature and evolution of high-z U/LIRGs to calculate the fraction of SNe missed as a function of redshift.

\cite{2012ApJ...756..111M} provide a parameterization of the form $k \times z+m$ for the missing SNe fraction as a function of redshift out to $z=2$. We extend this parameterization beyond $z=2$ by extrapolating the function beyond $z=2$. Using this model for the fraction of missed GRB OAs from host galaxy dust extinction, we obtain a missing GRB OA fraction of 30-35\%, which is comparable to recent estimates based on spectroscopy of GRB host galaxies. We note that this model assumes that GRB OAs are either detected (because of minimal host extinction) or not observed at all because of very high host extinction.

\subsection{Redshift desert}
\begin{figure}
\includegraphics[scale=0.68]{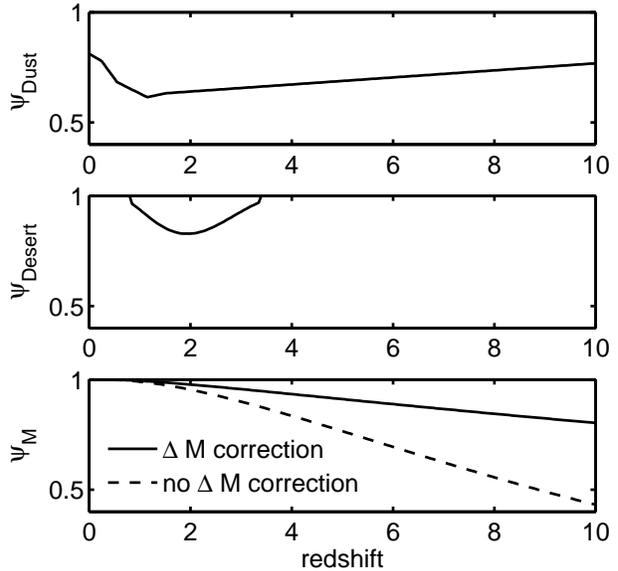}
\caption{{\bf Top panel:} The redshift selection function, $\psi_{\rm Dust}(z)$ is based on \citet{2012ApJ...756..111M}, who provide a parameterization for the missing SNe fraction as a function of redshift out to $z=2$. We extend this parameterization beyond $z=2$ by extrapolating the function values from $z=2-10$. {\bf Middle panel:} The curve shows the selection function for the redshift desert, based on a $85\%$ success rate of redshift measurement in galaxy surveys in $z=1.3-2.5$. {\bf Bottom panel:} The redshift selection function $\psi_{\rm M}(z)$ for two scenarios, the first, with $m_\mathrm{L}=24$, a mean time for redshift measurement $T_\mathrm{z}$ of 400 min, and a correction term for the $E_{{\rm iso}}-L_{{\rm optical}}$ correlation. The second scenario, dashed curve, is the same as above but without this correction. } \label{fig_dust}
\end{figure}

In addition to the effect of dust extinction, the redshift desert decreases the probability of acquiring a spectroscopic redshift in the $1.3 < z < 2.5$ interval. Quantifying this probability within this region is very difficult, because it depends on the strategy and brightness of the sources. As an observational constraint on the probability of redshift measurement in the desert we employ the success rate of acquiring galaxy spectroscopic redshifts from \cite{2005mmgf.conf..169S} i.e. $(70-80)\%$. The simplest functional form of $\psi_{\rm Desert}(z)$ is a smoothed step function with $\psi_{\rm Desert}(z)=80\%$ for $1.3 < z < 2.5$ and unity probability outside this region. Fig. \ref{fig_dust} plots the selection effect models $\psi_{\rm Dust}(z)$, $\psi_{\rm Desert}(z)$ and $\psi_{\rm M}(z)$. The total fraction missed over $z=0-10$ is less than 10\%.

\section{Results and discussion}
\begin{figure}
\centering
\includegraphics[scale=0.68]{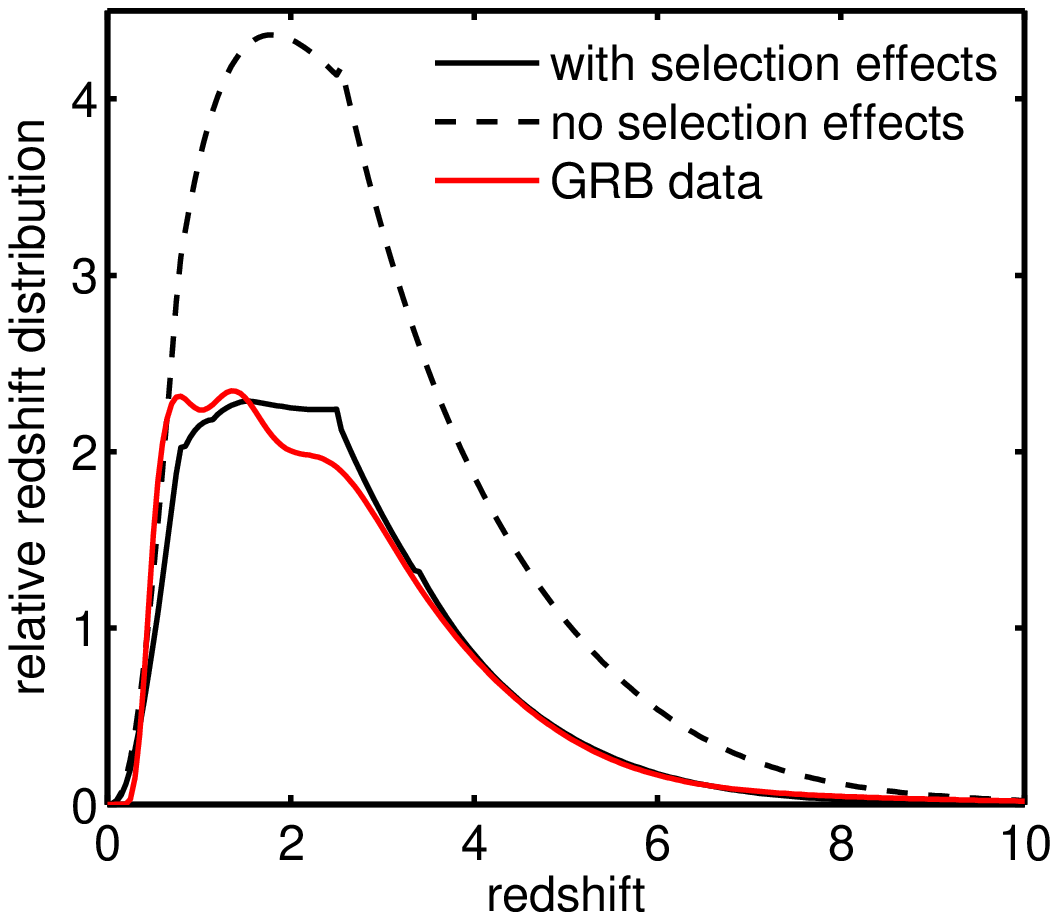}
\includegraphics[scale=0.68]{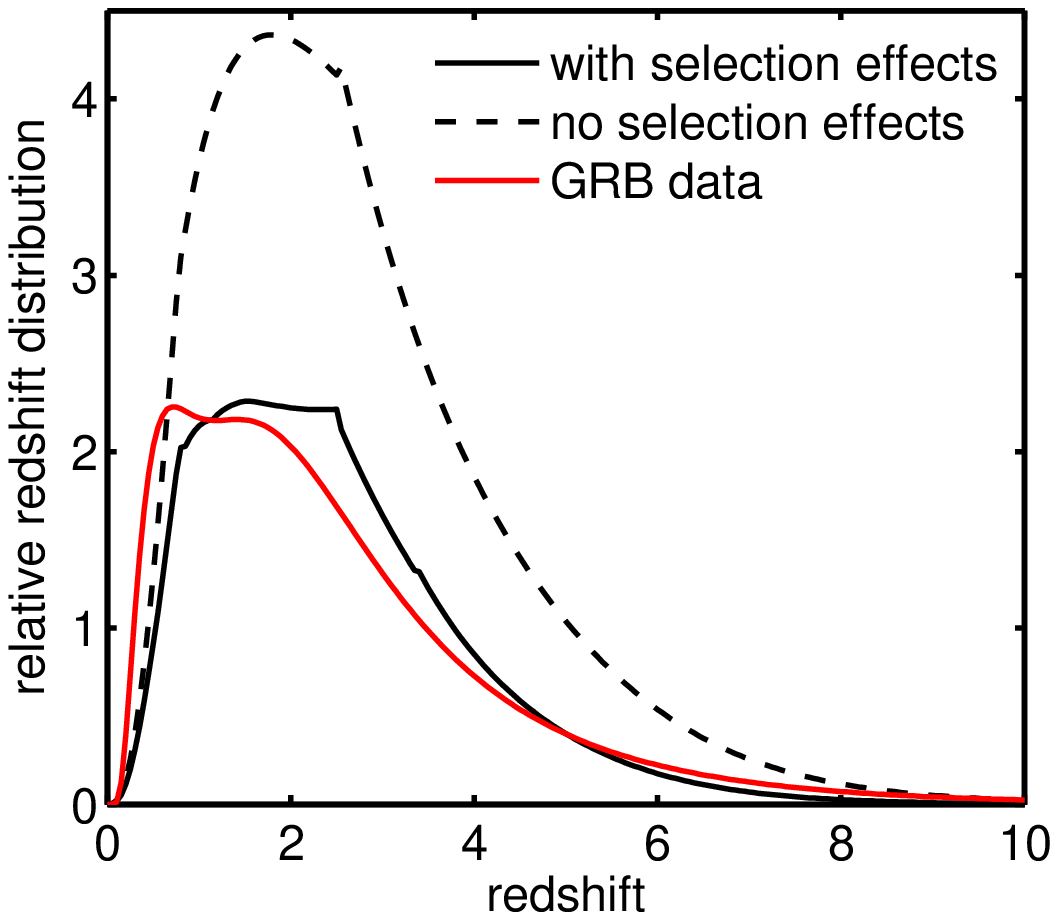}
\caption{{\bf Top:} The redshift distributions for models with and without optical selection effects, and the relative distribution of the HC sample. The optimal models, defined as having both high K-S probabilities and small fractional errors, are Models 1 and 3, both of which includes selection effects (shown in Fig. \ref{fig_dust}), a GRB rate evolution at $z=10$ similar to that of $z=1$, and either including or excluding a Malmquist bias correction. The least optimal model, Model 4, excludes selection effects. {\bf Bottom:} Same as the top figure but using the TOUGH redshift distribution. Both the HC and TOUGH data require the same optimal models (1 and 3).} \label{results}
\end{figure}

The inclusion of selection effects is required to achieve the maximum K-S probability, for the two data sets (HC and TOUGH), using the redshift cumulative distribution and model. Fig. \ref{results} plots the observed redshift distribution, with the optimal model (Model 1 below), and for comparison the expected distribution that would be observed if all optical selection effects were removed. Below we list the four model scenarios and Table 1 tabulates the K-S probabilities and fractional errors for the models:

\begin{enumerate}
\item  {\bf Model 1} -- Includes selection effects and the Malmquist bias correction term ($\Delta M(z)$), assuming $m_\mathrm{L}=24$ and $e(z=10)=1$, i.e. a SFR at $z=10$ similar to that at $z=0$.

\item {\bf Model 2} -- Includes selection effects, excludes Malmquist bias correction term, with $e(z=10)=4$, i.e. a SFR at $z=10$ four times higher than at $z=0$.

\item {\bf Model 3} -- Includes selection effects, excludes Malmquist bias correction term and uses the same GRB rate evolution as Model 1, $e(z=10)=1$.

\item {\bf Model 4} -- Excludes all optical selection effects. This model does not account for the missing redshifts from the  HC and TOUGH samples.
\end{enumerate}

\begin{table}\label{table1}
 \caption{The maximum KS statistics ($P_{\rm{KS}}$) using the model cumulative distributions for the HC and TOUGH samples. Models with selection effects (Models 1, 2 and 3) assume $m_\mathrm{L}=24$ and $T_z=400$ min. Also shown for each Model are the total fraction of unmeasured redshifts, $F_T$, and the error calculated from the {\it Swift} fraction of redshifts.}      
 \label{symbols}
 \begin{tabular}{@{}lcccccc}
  \hline
&  $P_{\rm{KS}}$ & $e(z=10)$$^\dagger$ & $\Delta M(z)$$^\ddagger$ & $F_{T}$ & $F_{T}$ error\%\\
   \hline
   {\bf HC} & & & & &\\
   Model 1 & 0.86 & 1  & yes & 65\%  & 3\% \\
   Model 2 & 0.27 & 4  & yes & 71\%  & 13\%\\
   Model 3 & 0.96 & 1  & no & 70\%  & 11\% \\
   Model 4 & 0.13 & 1  & no & 0\%  & 100\%\\
  \hline
  {\bf TOUGH} & & & & &\\
    Model 1 & 0.68 & 1  & yes & 47\%  & 4\% \\
   Model 2 & 0.17 & 4  & yes & 53\%  & 18\%\\
   Model 3 & 0.86 & 1  & no & 52\%  & 16\% \\
   Model 4 & 0.15 & 1 & no & 0\%  & 100\%\\
   \hline
   \hline
\end{tabular}
 \smallskip
 $^\dagger$ GRB rate evolution $e(z=10)=1$ is an extrapolation from the fit of \citet{2009ApJ...692..778R}.
$^\ddagger$ the Malmquist correction is $\Delta M(z)\approx 1.7 \rm{Log}_{10}(1+z)$.
\end{table}

Even though model 4 (which excludes all optical selection effects) is not rejected by the KS test ($P=13$\%), it does not account in any way for the missing 45\% and 63\% (TOUGH and HC data) of redshifts from the {\it Swift} sample. Alternatively, Model 1, including realistic selection effects, is compatible with the observed distribution including a GRB rate evolution that follows the global SFR. 

It is useful to stress how different selection effects could have similar affects on the redshift  distribution. For example, there are different biases that could potentially reproduce excess number counts at high-$z$:
\begin{enumerate}
\item A larger intrinsic GRB rate at very high-$z$ ($z=10$) via SFR or metallicity evolution compared to the SFR at $z=0$ (Model 2).
\item Less dust obscuration of GRB hosts at high-$z$, resulting in brighter OAs.
\item GRB circumburst environments could be relatively different at high-$z$, which would also affect the OA brightness.
\item Correlation between high $E_{{\rm iso}}$ and $L_{{\rm opt}}$ bursts boost the fraction of redshifts measurable at high-$z$.
\item An evolving GRB high energy LF. Given the correlation between $E_{{\rm iso}}$ and $L_{{\rm opt}}$, an evolving high energy luminosity will further increase the optical luminosities at high-$z$, so that redshift measurement is more probable.
\end{enumerate}

In summary, our analysis suggests that a combination of selection effects (both instrumental and astrophysical) can adequately describe the observed redshift distribution. Furthermore the observed distribution is compatible with a rate evolution that tracks the evolving SFR. We have also highlighted that two GRB redshift catalogues claiming to be statistically complete, are in fact not compatible with each other. We show that the TOUGH selection and a subset of absorption redshifts (the HC sub-sample) are compatible.

The optimal strategy to avoid the complicated selection biases in the current redshift sample is to employ dedicated ground based instruments with well understood sensitivities and performance. Robotic instruments with integral field low resolution spectroscopic capabilities could also play an important role, both in terms of constraining redshifts and studying the early optical emissions. Also, high-energy detection combined with simultaneous optical coverage will continue to play a role in increasing the number of redshifts, especially when combined with a dedicated ground-based NIR follow-up network. 

%



\section*{Acknowledgments}
D.M. Coward is supported by an Australian Research Council Future Fellowship. E.J. Howell acknowledges support from a University of Western Australia Fellowship.

\appendix

\section[]{The {\it Swift} GRB selection function}\label{app}
The GRB selection function is defined as the fraction of potentially observable bursts by {\it Swift} as a function of redshift. It depends on the LF of the bursts and the sensitivity of the BAT detector to GRBs. To account for the flux limit of the {\it Swift} BAT detector we include two corrections: the cosmological k-correction, and $B_{det}$, the bolometric correction of the detector:
\begin{equation}
B_{det} \equiv  \frac{\int_{1{\rm keV}}^{10000{\rm keV}} E N(E) {\rm d}E}{\int_{e_2}^{e_1} E N(E) {\rm d}E}\;,
\end{equation}
where the interval $[e_1, e_2]$ is the spectral energy band that the detector is sensitive to i.e $(150-15)$keV and $N(E)$ is the ensemble photon flux density ($\mathrm {ph\;s^{-1}\;cm^{-2}\;keV^{-1}}$) at the photon energy $E$. The model burst spectrum we employ is the Band function \citep{band03}, with $\alpha = -1$ and $\beta = -2.3$ and a rest frame break energy of 511 keV. We employ a k-correction of the form:
\begin{equation}
k(z) \equiv  \frac{\int_{15{\rm keV}}^{150{\rm keV}} E N(E) {\rm d}E}{\int_{(1+z)15{\rm keV}}^{(1+z)150{\rm keV}} E N(E) {\rm d}E}\;.
\end{equation}
 
For the GRB high energy LF we use the single power law form used by \citet{pm01} which has an exponential cutoff at low luminosity:
\vspace{-0.5mm}
\begin{equation}
\Phi(L) = \Phi_{0}\hspace{1.0mm}\left(\frac{L}{L_{*}}\right) ^{-\alpha} \mathrm{exp}\left( -\frac{L_{*}}{L}\right)\,.
\end{equation}
 
\noindent Here, $L$ is the peak rest frame photon luminosity in the 1-10000\,keV energy range, $\alpha$ ensures an asymptotic slope at the bright end and $L_{*}$ is a characteristic cutoff scaling. This form was used by \citet{2013MNRAS.428..167H} to perform $\chi^{2}$ minimisation of the \emph{Swift} LGRBs brightness distribution. We adopt their best fitting parameters of: $L_{*} =2.0^{+0.2}_{-0.02}\times 10^{52}\, \mathrm{erg}\,\mathrm{s}^{-1}
\mathrm{and}\,\, \alpha =3.8 ^{+0.2}_{-0.6}$.
 
The fraction of detectable GRBs, or flux-limited selection function (those observed with peak flux  $>F_{\rm lim}$) is
\begin{equation}\label{grbrate}
\psi_{\rm Swift}(z)= \int_{L_{\rm lim}(F_{\rm lim},z)}^{L_{\rm max}}\phi(L) {\rm d}L \;,
\end{equation}
where the luminosity is an isotropic equivalent luminosity and $L_{\rm lim}(F_{\rm lim},z)$ is obtained by solving
\begin{equation}
F_{\rm lim}(L_{\rm lim},z) = \frac{L_{\rm lim}} {4 \pi d^2_{\mathrm L}(z)}k(z) B_{det},
\end{equation}
where $d_{\mathrm L}(z)$ denotes the luminosity distance of the burst using a flat-$\Lambda$ cosmology with $H_{\mathrm 0}$ = 71 km s$^{-1}$ Mpc$^{-1}$, $\Omega_M$ = 0.3 and $\Omega_\Lambda$ = 0.7. The flux limit, $F_{\rm lim}$, for GRB detection of {\it Swift}, we use the smallest peak flux of $0.04$ ph cm$^{-2}$ s$^{-1}$ from the \cite{2013MNRAS.428..167H} sample.

\section{GRB rate evolution}\label{ez}
\begin{figure}
\includegraphics[scale=0.6]{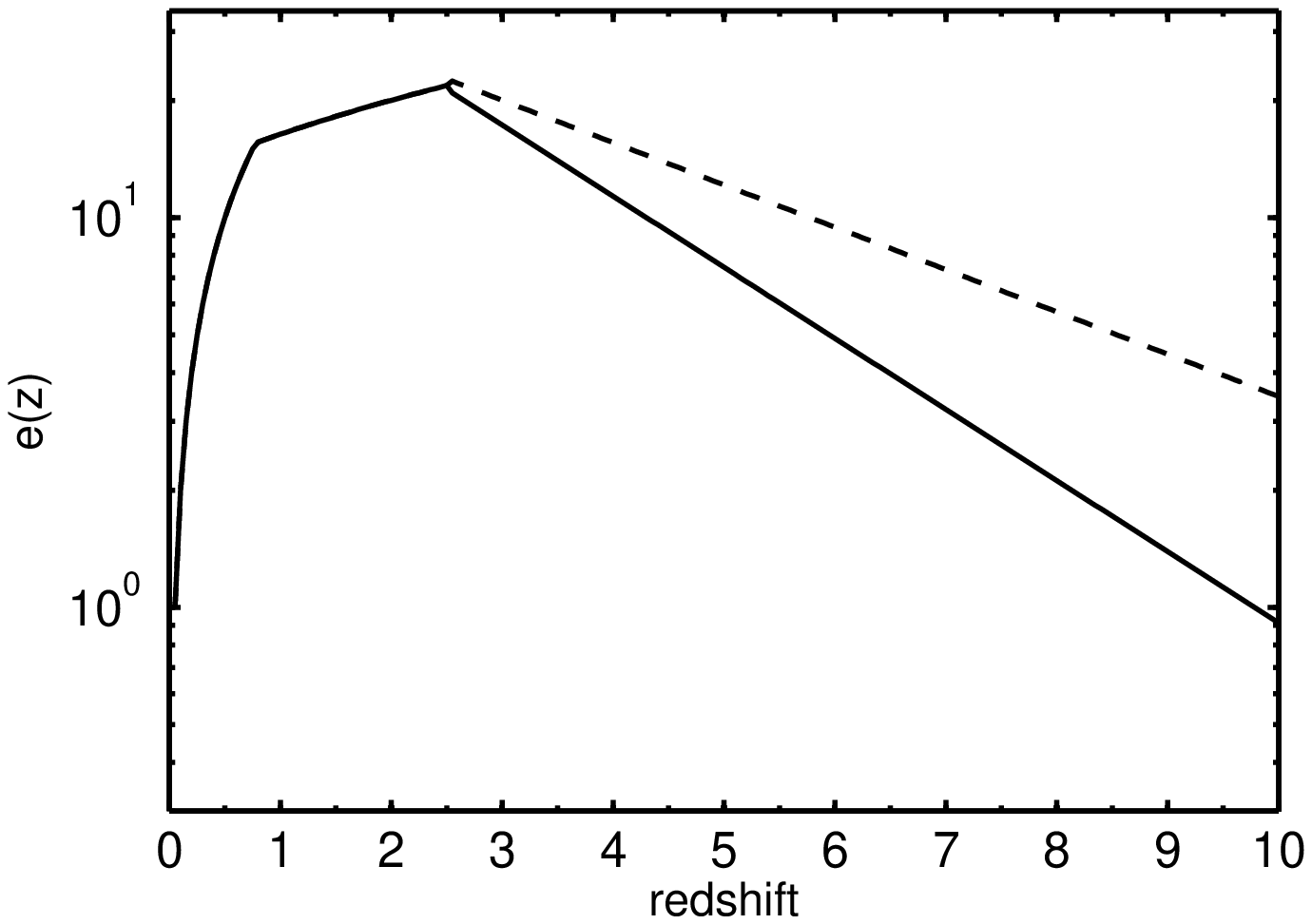}
\caption{Assuming that GRB rate evolution tracks the SFR, we extrapolate the fit from \citet{2009ApJ...692..778R} to $z=10$, using equation (\ref{ezeq}). We convert this fit to a dimensionless evolution function, $e(z) = $SFRD$(z)/$SFRD$(0)$. The solid curve is the extrapolation to $z=10$ used in Model 1, and the dashed curve is used to produce Model 2 (see Appendix \ref{ez} for the parameters used for $e(z)$).} \label{ezfig}
\end{figure}

Assuming GRB rate evolution tracks the SFR, we use the following function to approximate GRB rate evolution:
\begin{equation}\label{ezeq}
  e(z)=\biggl \lbrace{ \begin{array}{lll}
                   \vspace{1mm}11z & z < 0.8 \\
                   \vspace{1mm}7+2z & 0.8 \leq z < 2.5 \\
                   a[\mathrm{exp}(z-3)]^{-b} & z > 2.5
                 \end{array}}
\end{equation}
For the optimal GRB redshift distribution model plotted in Fig. \ref{results} we employ $a=9.5$ and $b=0.42$. For the model excluding OA brightness evolution, the optimal model requires $e(z)$ parameters $a=11$ and $b=0.25$, yielding a GRB rate evolution about 4 times higher at $z=10$ compared to the present day (see Fig. \ref{ezfig}).

\label{lastpage}
\end{document}